\documentclass[letter]{aa}     

\usepackage{graphicx}
\usepackage[varg]{txfonts}
\usepackage{epstopdf}
\usepackage{lipsum}
\usepackage{subcaption}         
\usepackage{lscape}             
\usepackage{placeins}           
                                
\usepackage{amsmath}

\usepackage[colorlinks=true,linkcolor=magenta,citecolor=blue,filecolor=black,urlcolor=cyan,final=true]{hyperref}

\begin{document}

   \title{Observing the solar corona from a formation-flying mission}

   \subtitle{First results of Proba-3/ASPIICS}

   \author{
   A.~N.~Zhukov\inst{\ref{ROB}, \ref{SINP}}
   \and
   L.~Dolla\inst{\ref{ROB}}
   \and
   M.~Mierla\inst{\ref{ROB},\ref{RomanianAcademy}}
   \and
   B.~D.~Patel\inst{\ref{ROB}}
   \and
   S.~Shestov\inst{\ref{ROB},\ref{CSL}}
   \and
   B.~Bourgoignie\inst{\ref{ROB}}
   \and
   A.~Debrabandere\inst{\ref{ROB}}
   \and
   C.~Jean\inst{\ref{ROB}}
   \and
   B.~Nicula\inst{\ref{ROB}}
   \and
   D.-C.~Talpeanu\inst{\ref{ROB}}
   \and
    Z.~Zontou\inst{\ref{ROB}}
   \and
   S.~Fineschi\inst{\ref{OATo}}
    \and
   S.~Gun\'ar\inst{\ref{CAS}}
    \and
   P.~Lamy\inst{\ref{LATMOS}}
   \and
   H.~Peter\inst{\ref{MPS}}
    \and
   P.~Rudawy\inst{\ref{UWroclaw}}
    \and
   K.~Tsinganos\inst{\ref{UoA}}
    \and
   L.~Abbo\inst{\ref{OATo}}
   \and
   C.~Aime\inst{\ref{UNice}}
   \and
   F.~Auch\`ere\inst{\ref{IAS}}
   \and
   D.~Berghmans\inst{\ref{ROB}}
   \and
   D.~Be\k{s}liu-Ionescu\inst{\ref{AIRA}, \ref{RomanianAcademy}}
   \and
   S.~E.~Gibson\inst{\ref{HAO}}
   \and
   S.~Giordano\inst{\ref{OATo}}
   \and
   P.~Heinzel\inst{\ref{CAS},\ref{UWroclaw}}
   \and
   B.~Inhester\inst{\ref{MPS}}
   \and
   J.~Magdaleni\'c\inst{\ref{ROB}, \ref{KUL}}
   \and
   C.~Marqu\'e\inst{\ref{ROB}}
   \and
   L.~Rodriguez\inst{\ref{ROB}}
   \and
   M.~St\k{e}\'slicki\inst{\ref{CBK}}
   \and
   L.~Zangrilli\inst{\ref{OATo}}
   \and
   D.~Galano\inst{\ref{ESTEC}}
   \and
   R.~Rougeot\inst{\ref{ESTEC}}
   \and
   J.~Versluys\inst{\ref{ESTEC}}
   \and
   C.~Thizy\inst{\ref{CSL}}
     }

  \institute{
  Solar--Terrestrial Centre of Excellence --- SIDC, Royal Observatory of Belgium, 1180 Brussels, Belgium\\
  \email{Andrei.Zhukov@sidc.be}
  \label{ROB}
   \and
  Skobeltsyn Institute of Nuclear Physics, Moscow State University, 119991 Moscow, Russia
  \label{SINP}
    \and
  Institute of Geodynamics of the Romanian Academy, 020032 Bucharest-37, Romania
  \label{RomanianAcademy}
  \and 
  Centre Spatial de Li\`ege, Universit\'e de Li\`ege, Av. du Pr\'e-Aily B29, 4031 Angleur, Belgium
  \label{CSL}
 \and
  National Institute for Astrophysics, Astrophysical Observatory of Torino, Pino Torinese, Torino, Italy
  \label{OATo}
 \and
  Astronomical Institute of the Czech Academy of Sciences, 251 65 Ond\v{r}ejov, Czech Republic
  \label{CAS}
  \and
  Laboratoire Atmosph\`eres et Observations Spatiales, 11 Boulevard d’Alembert, 78280 Guyancourt, France
  \label{LATMOS}
 \and
            Max Planck Institute for Solar System Research, Justus-von-Liebig-Weg 3, 37077 G\"ottingen, Germany\label{MPS}     
  \and
  Astronomical Institute, University of Wroc\l{}aw, Kopernika 11, 51-622 Wroc\l{}aw, Poland
  \label{UWroclaw}
  \and
  University of Athens, Panepistimiopolis, 157 84 Zografos Athens, Greece
  \label{UoA}
  \and
  Universit\'e C\^{o}te d’Azur, CNRS, Observatoire de la C\^{o}te d’Azur, UMR7293 Lagrange,
Parc Valrose, 06108, Nice, France
  \label{UNice}
  \and
  Institut d'Astrophysique Spatiale, Orsay, France
  \label{IAS}
    \and
  Astronomical Institute of the Romanian Academy, Bucharest, Romania
  \label{AIRA}
  \and
  High Altitude Observatory, National Center for Atmospheric Research, Boulder, CO, USA
  \label{HAO}
  \and
  Center for mathematical Plasma Astrophysics, KU Leuven, Leuven 3000, Belgium
  \label{KUL}
 \and
  Centrum Bada\'n Kosmicznych Polskiej Akademii Nauk, Poland
  \label{CBK}
   \and
  European Space Research and Technology Centre, European Space Agency, Noordwijk, Netherlands
  \label{ESTEC}
             }

   \date{Received / Accepted}

 
  \abstract
  {
   We report the first results from observations of the solar corona by the ASPIICS coronagraph aboard the \mbox{Proba-3} mission. ASPIICS (Association of Spacecraft for Polarimetric and Imaging Investigation of the Corona of the Sun) is a giant coronagraph consisting of the telescope mounted aboard one of the mission's spacecraft and the external occulter placed on the second spacecraft. The two spacecraft separated by around 144~m fly in a precise formation up to 5.5 hours at a time, which allows coronal observations in eclipse-like conditions, i.e. close to the limb (typically down to 1.099~$R_\odot$, occasionally down to 1.05~$R_\odot$) and with very low straylight. 
ASPIICS observes quasi-stationary structures, such as coronal loops, streamers, quiescent prominences, and a variety of dynamic phenomena: erupting prominences, coronal mass ejections, jets, slow solar wind outflows, coronal inflows. In particular, weak, widespread and persistent small-scale outflows and inflows between 1.3 and 3~$R_\odot$ are observed at a high spatial (5\farcs6) and temporal (30~s) resolution for the first time, expanding the range of scales at which the variable slow solar wind is observed to form. 
}


   \maketitle
%

\section{Introduction}
\label{S-intro}

Understanding the physics of the low and middle solar corona (from the coronal base up to a few solar radii $R_\odot$) is important for solving problems such as the origin of the variable slow solar wind \citep[e.g.][]{Abbo2016} and the physics of coronal mass ejections \citep[CMEs, see e.g.][]{Howard2023}. Observing the corona between around 1.2 and 2.5~$R_\odot$ is, however, difficult as this region lies at the interface between the fields of view of extreme-ultraviolet (EUV) imagers and traditional externally occulted coronagraphs, see e.g. a review by \citet{West2023}. Innovative observations by the ASPIICS coronagraph (Association of Spacecraft for Polarimetric and Imaging Investigation of the Corona of the Sun) aboard the formation flying Proba-3 mission of the European Space Agency (ESA), aim to fill in this observational gap \citep{Zhukov2025}. Proba-3 consists of two spacecraft: one carries the optical telescope and the other carries the external occulter of the coronagraph. The large distance between the two spacecraft (around 144~m) and the millimetric precision of their positioning with respect to each other allow uninterrupted observations of the corona for up to 5.5~hours and with very low straylight, effectively emulating conditions encountered during a total eclipse. In this Letter, we report the first results from coronal observations by Proba-3/ASPIICS. 

\begin{figure*}[t!]
\centering
\begin{subfigure}[t]{0.46\textwidth}
\centering
\includegraphics[width=1.0\textwidth]{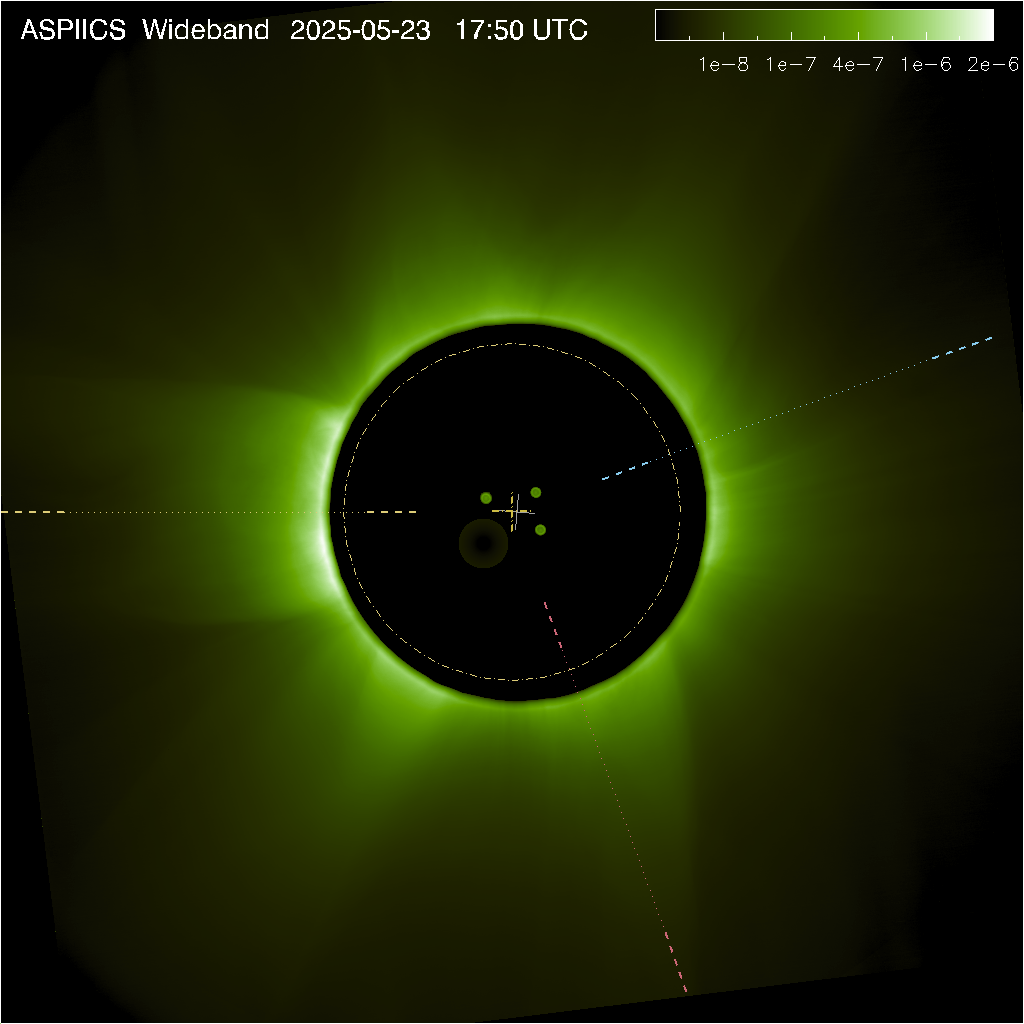}
\end{subfigure}~\begin{subfigure}[t]{0.5\textwidth}
\centering
\includegraphics[width=1.0\textwidth]{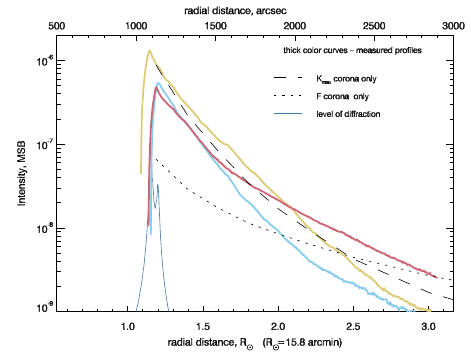}
\end{subfigure}
\caption{Left panel: the solar K-corona observed by Proba-3/ASPIICS on 23 May 2025 at 17:50:31~UTC (orbit 207) in the wide (about 300~\AA) spectral passband centered at 5510~\AA. This high dynamic range image is composed of three images taken with exposure times of 0.1~s, 1~s, and 10~s. The internal occulter center is plotted as a gray cross. The  position of the solar center and the solar disk are shown as the yellow cross and circle, respectively. Three dotted lines mark the directions along which the brightness is plotted in the right panel: dark region with seemingly open magnetic field (blue), bright streamer (red), and post-CME streamer (yellow). Three dots in the middle are the LEDs of the Occulter Position Sensor Emitter \citep{Zhukov2025}, additional weak structures in the middle are due to the light reflections on the  backside of the occulter. In all the images shown in this Letter solar north is up, west is to the right. Right panel: profiles of the K-corona brightness along the dotted lines indicated in the left panel, plotted as a function of radial distance in the corresponding color. The brightness turnover around 1.1~$R_\odot$ is due to the instrumental vignetting. The dashed and dotted lines show the typical intensities of the K-corona during solar maximum and F-corona, respectively \citep{Cox2000}. The thin solid line shows the diffracted light intensity calculated using the model by \citet{Shestov2018}.   
}
\label{fig:radiometry}
\end{figure*}


\section{Observations}
\label{S-observations}

After its launch on 5 December 2024 and commissioning, \mbox{Proba-3} entered its nominal mission phase on 3 July 2025. The ASPIICS telescope and its general operational procedures are described by \citet{Zhukov2025}. To cover the large dynamic range of brightness in each coronal scene, three images with different exposure times are taken in a quick succession. The cadence between successive triplets (``acquisitions'') is up to 30~s. Each image is first radiometrically calibrated. The calibration parameters obtained during the on-ground calibration campaign \citep{Fineschi2025, Zhukov2025} are used\footnote{The calibration parameters were verified using observations of stars during the in-flight calibration campaign on 20 December 2024.}. The resulting three Level-2 images calibrated in MSB (Mean Solar Brightness) are then aligned with respect to the solar center and merged to obtain a Level-3 image with high dynamic range (HDR) in brightness, covering the whole field of view up to 3~$R_\odot$. The F-corona was subtracted using the model by \citet{Cox2000}. 

\begin{figure}[ht]
\centering
\includegraphics[width=0.4\textwidth]{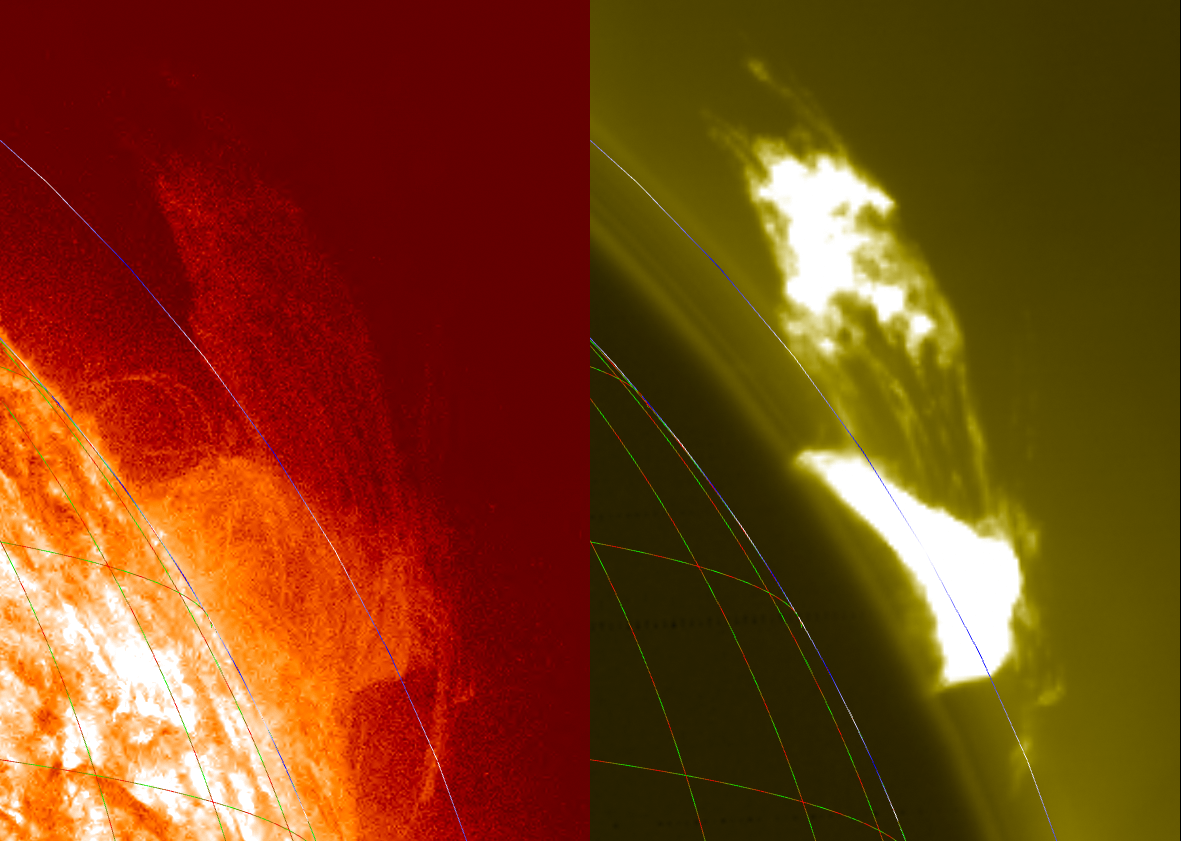}
\caption{Left panel: a quiescent prominence observed by SDO/AIA in the He~II passband (304~\AA) on 25 March 2025 at 19:23:29~UTC. Right panel: a co-aligned image of the same prominence taken by ASPIICS in the He~I D$_3$ passband (5876~\AA) on 25 March 2025 at 19:23:23~UTC (orbit 135). In both panels, the solar coordinate grid is shown as well as the nominal inner edge of the ASPIICS field of view at 1.099~$R_\odot$.}
\label{fig:he}
\end{figure}

\section{Results}
\label{S-results}

\subsection{Imaging the quiescent corona}
\label{S-quiescent}

\begin{figure*}
\sidecaption
\includegraphics[width=12cm]{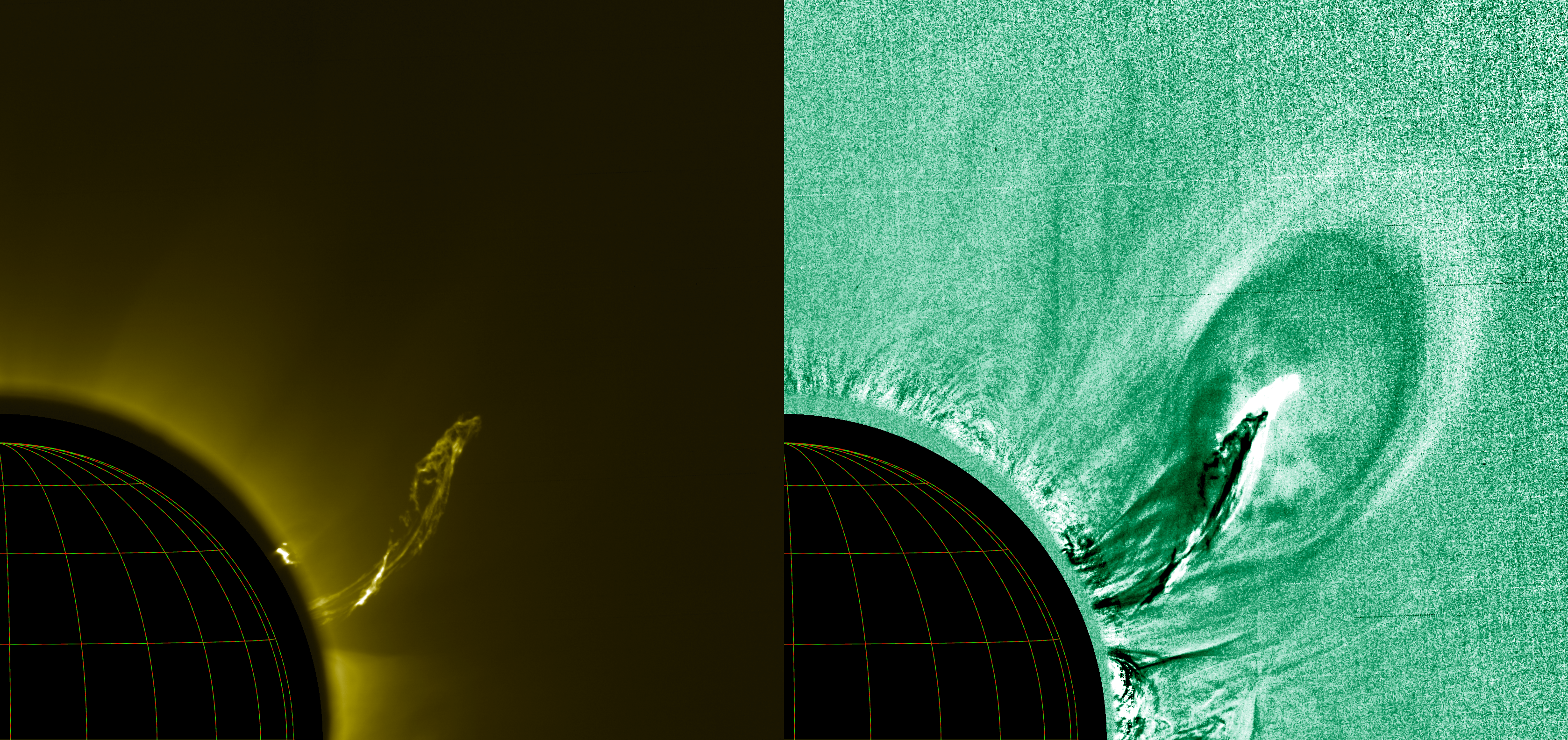}
\caption{A CME observed by Proba-3/ASPIICS on 16 June 2025 (orbit 236). Left panel: He~I D$_3$ (5876~\AA) image of the corresponding erupting prominence taken at 07:27:41~UTC. The prominence twist is clearly visible. Right panel: Fe~XIV (5303~\AA) image of the CME taken at 07:38:47~UTC, with the image taken at 07:28:58~UTC subtracted. The CME bright front, cavity, and the erupting prominence can be seen. A radial filter was applied to enhance the visibility of the weak bright front. The associated movie (Movie 1) is available online.}
\label{fig:CME}
\end{figure*}

Figure~\ref{fig:radiometry} shows an image of the quiescent corona taken on 23 May 2025 during the 207th orbit of Proba-3 since the launch\footnote{Coronal observations by Proba-3 last up to 5.5~hours around the orbit apogee, which can extend over two calendar days. It is convenient to identify the coronagraphy windows by the number of the orbit.}. The image reveals the large-scale coronal structure typical for the solar cycle maximum, with streamers around the entire disk. The calibrated coronal brightness (Figure~\ref{fig:radiometry}, right panel) is consistent with previous measurements of the maximum-type corona. The impact of the diffracted light remains to be analyzed in detail, but it appears to be small. Applying an image enhancement algorithm reveals the coronal fine structure throughout the field of view (Figure~\ref{fig:wb_wow}). 

The nominal inner edge of the ASPIICS field of view is at 1.099~$R_\odot$ (start of 100\% vignetting). Figure~\ref{fig:he} demonstrates that due to  jitter in the coronagraph spacecraft's pointing the inner edge of the field of view can be occasionally as low as 1.05~$R_\odot$ (with the corresponding simultaneous increase of the inner edge position on the opposite side of the Sun). The prominence fine structure seen in the right panel of Figure~\ref{fig:he} (down to the width of a couple of pixels) demonstrates that the ASPIICS spatial resolution is close to the theoretical limit of two pixels (around 5\farcs6).

ASPIICS has three polarizers mounted at 60$^\circ$ with respect to each other to determine coronal polarization (Figure~\ref{fig:polar_image}). Dividing the polarized brightness by the total brightness obtained from unpolarized measurements (Figure~\ref{fig:radiometry}), we obtain the polarization degree of the corona, which is in reasonable agreement with theoretical expectations (Figure~\ref{fig:polar_plots}, bottom panel). The polarization tangential to the solar limb is recovered with a precision of around 2$^\circ$ (Figure~\ref{fig:density}), which may be due to an inaccurate determination of the position of the solar disk. The coronal electron number density can be inverted from the polarized brightness measurements using the method by \citet{vandeHulst1950}, giving values that are in agreement with previous observations (Figure~\ref{fig:density}). 

Coronal loops observed by ASPIICS in the Fe~XIV passband (5303~\AA) appear remarkably similar to the loops observed in the Fe~XIV passband (211~\AA) by the Atmospheric Imaging Assembly \citep[AIA, ][]{Lemen2012} aboard the Solar Dynamics Observatory (SDO). Figure~\ref{fig:fe} shows that hot 2~MK loops are best seen in Fe~XIV by both ASPIICS and AIA, whereas the ASPIICS wideband image contains structures at various temperatures integrated along the line of sight, leading to low contrast for these hot loops. 

\subsection{Coronal mass ejection}
\label{S-CME}

One of the first CMEs observed by ASPIICS is shown in Figure~\ref{fig:CME}. The CME was already well under way when ASPIICS observations started at 07:16~UTC, with the typical three-part CME structure (bright front, cavity, and erupting prominence) clearly visible in a few images taken in all passbands. Starting from 07:28~UTC, the ASPIICS filter wheel got temporarily stuck near the Fe~XIV position, so most of the ASPIICS measurements of the CME kinematics were made in the Fe~XIV passband. 

\begin{figure}[!ht]
\centering
\includegraphics[width=0.49\textwidth]{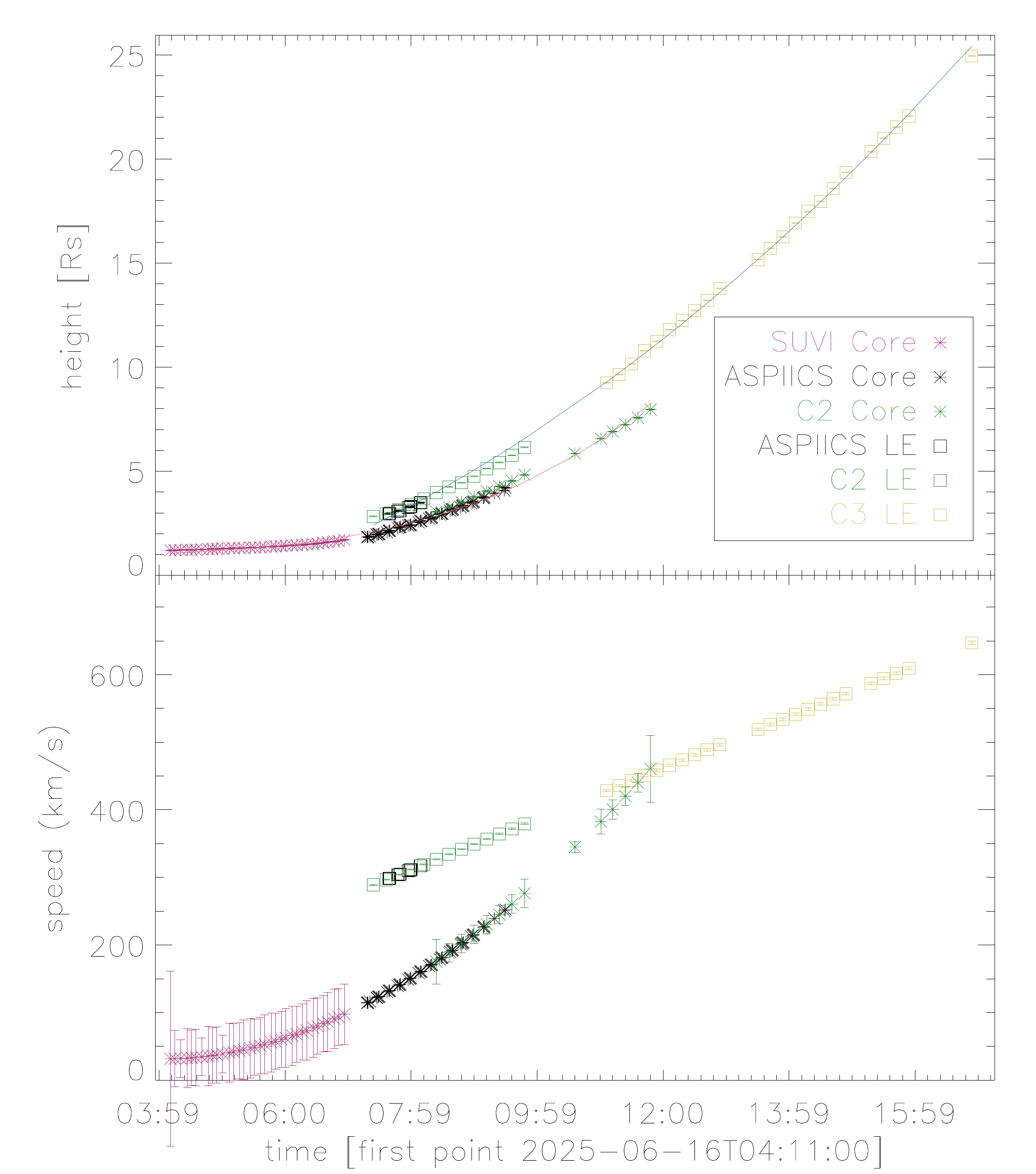}
\caption{Kinematics of a CME observed by Proba-3/ASPIICS on 16 June 2025 (orbit 236), GOES/SUVI, and SOHO/LASCO C2 and C3. Top and bottom panels show the evolution of the height and radial velocity. Asterisks and squares show the propagation of the CME core and the leading edge (LE) respectively. A linear plus power law fit for the CME core propagation is shown as a purple line. A quadratic fit for the leading edge propagation is shown as a blue line. The velocity was calculated using the three-point differentiation. 
}
\label{fig:kinematics}
\end{figure}

Figure~\ref{fig:kinematics} shows the evolution of the CME height and speed as measured by ASPIICS, Solar UltraViolet Imager \citep[SUVI, see ][]{Tadikonda2019} of a Geostationary Operational Environmental Satellite (GOES), and LASCO \citep[Large-scale Spectroscopic Coronagraph, see][]{Brueckner1995} aboard the Solar and Heliospheric Observatory (SOHO). ASPIICS was the only instrument to measure the propagation of the CME core between 1.5~$R_\odot$ and 3~$R_\odot$. The CME core kinematics in this case is best fit with a sum of a linear function dominant during the initial slow rise and a power law governing the fast acceleration phase later on \citep{Cheng2020}. This indicates that the initial perturbation that triggered the CME was sizable \citep{Schrijver2008}. 

\subsection{Small-scale dynamics in the slow solar wind formation region}
\label{S-wind}

\begin{figure}[!ht]
\centering
\includegraphics[width=0.49\textwidth]{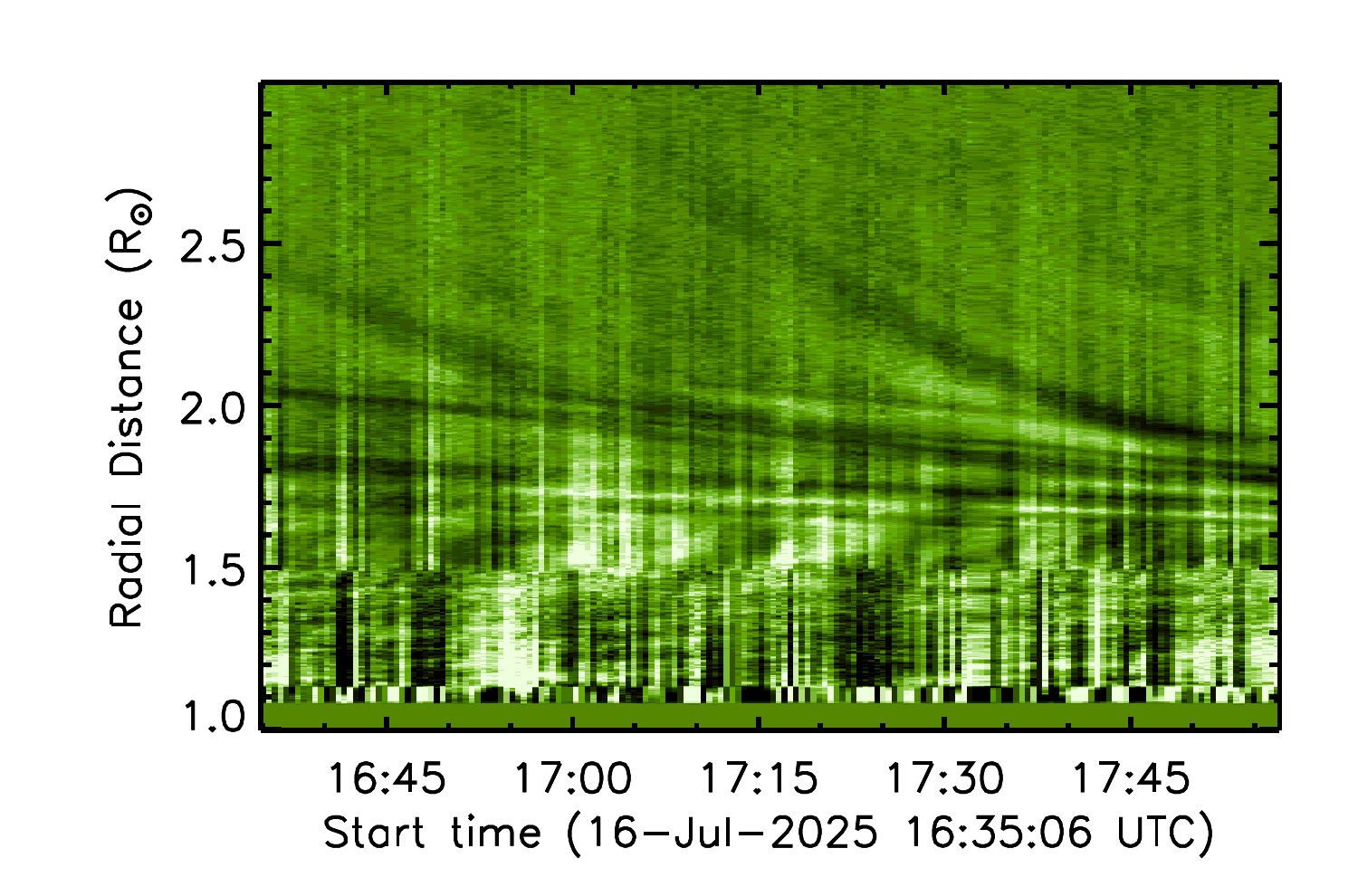}
\caption{Distance-time map of the radial propagation of small-scale structures at the position angle of 204$^\circ$. The coronal brightness was averaged over the 2$^\circ$-wide sector (see Figure~\ref{fig:track}). An image taken 5 minutes earlier was subtracted from every image. The cadence of the data was 30~s.}
\label{fig:jmap}
\end{figure}

ASPIICS reveals a rich variety of weak small-scale dynamics in the solar wind formation region. Both outflows and inflows are observed (Figure~\ref{fig:jmap}), sometimes overlapping in projection at the same location (Figure~\ref{fig:track}). Numerous narrow jets are also observed (Figure~\ref{fig:jet}), in some cases repeatedly originating from the same location. Such jets have been shown to be the result of the emergence of magnetic flux into the solar atmosphere and its interaction with a pre-existing magnetic field of an active region \citep[e.g.][]{Archontis2010}. However, most of the bright outflows are similar to blobs \citep{Sheeley1997}, with their spatial scale (down to a few thousand km) significantly smaller than that of blobs in LASCO C2 observations. They are also much more frequent (around ten outflows can be distinguished in a narrow sector shown in Figure~\ref{fig:jmap} during one hour of observations) and are seen essentially at all position angles. Their speeds are between 100~km~s$^{-1}$ and 400~km~s$^{-1}$, without a clear acceleration. 

Coronal inflows, first reported in LASCO observations and interpreted as  resulting from reconnection at the folds of the heliospheric current sheet \citep{Sheeley2001}, are also abundant in the ASPIICS field of view. The inflows are usually dark and propagate at speeds around 100~km~s$^{-1}$, although two inflows most prominent in Figure~\ref{fig:jmap} decelerate from 350~km~s$^{-1}$ to around 100~km~s$^{-1}$, descending from around 3~$R_\odot$ to less than 2~$R_\odot$ in around an hour. Both inflows and outflows are probably a part of the S-web dynamics \citep{Antiochos2011} in the solar turbosphere, the region where large-scale solar wind outflow is not yet dominant and both small-scale inflows and outflows may co-exist \citep{Veselovsky1999}. The omnipresent and complex fine-scale dynamics is observed so clearly in this region of the corona for the first time. 

\section{Conclusions}
\label{S-discussion}

The ASPIICS data demonstrate that two spacecraft flying in formation can be used for coronagraphic observations of the inner solar corona. ASPIICS typically observes the corona as close to the solar center as 1.099~$R_\odot$, (occasionally down to 1.05~$R_\odot$), and out to 3~$R_\odot$. ASPIICS can image the structure of the solar corona both morphologically (Figure~\ref{fig:wb_wow}) and photometrically (Figure~\ref{fig:radiometry}). CME observations demonstrate that the ASPIICS data can fill the gap between the low corona (typically observed by EUV imagers like SUVI) and the high corona (typically observed by externally occulted coronagraphs like LASCO C2), where observations are difficult. A rich variety of omnipresent and dynamic fine structure (jets, outflows, inflows) is reported in this height range for the first time. It is probably produced by dynamics in the S-web and may shed light on the processes responsible for the origin of the variable slow solar wind.


\begin{acknowledgements}
      {The ASPIICS data are courtesy of the Proba-3/ASPIICS consortium. Proba-3 is a technology demonstration mission of the European Space Agency (ESA) and a Mission of Opportunity in the ESA Science Programme. The ASPIICS project is developed under the ESA’s General Support Technology Programme (GSTP) and the ESA’s PRODEX Programme thanks to the contributions of Belgium, Poland, Romania, Italy, Ireland, Greece, and the Czech Republic. The ROB team thanks the Belgian Federal Science Policy Office (BELSPO) for the provision of financial support in the framework of the PRODEX Programme of ESA under contract numbers 4000117262, 4000134474, 4000136424, 4000145189, and 4000147286. S.E.G. acknowledges support from NASA grant 80NSSC21K1860 (US Co-Is of the ASPIICS coronagraph). S.Gun\'ar and P.H. acknowledge the support from grant 25-18282S of the Czech Science Foundation and from the project RVO:67985815 of the Astronomical Institute of the Czech Academy of Sciences. P.H. was partially supported by the program ``Excellence Initiative - Research University'' for the years 2020--2026 for the University of Wroc\l{}aw, project No. BPIDUB.4610.96.2021.KG. M.S. is supported by grant No. 2024/55/B/ST9/03199 of the National Science Centre, Poland.}
\end{acknowledgements}

\bibliographystyle{aa}
\bibliography{bibliography.bib}

\begin{appendix}
\section{Supplementary figures}


\begin{figure*}[h!]
\centering
\includegraphics[width = 1.0\textwidth]{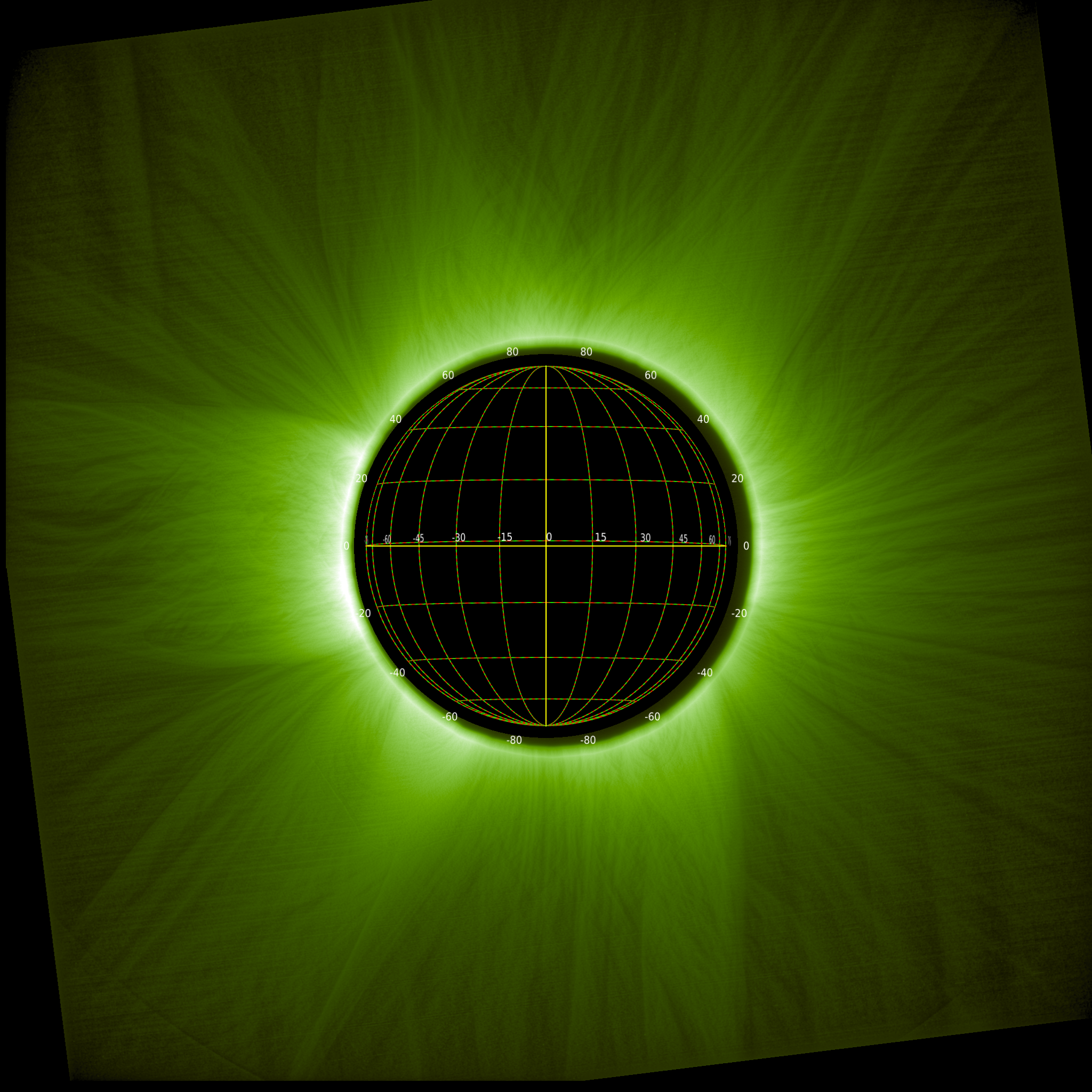}
\caption{The solar corona observed by Proba-3/ASPIICS on 23 May 2025 at 17:50:31~UTC (orbit 207) in the wide spectral passband centered at 5510~\AA\, (same image as in Figure~\ref{fig:radiometry}), but with the Wavelet-Optimized Whitening (WOW) algorithm \citep{Auchere2023WOW} applied to enhance the visibility of the fine structure of the corona. 
}
\label{fig:wb_wow}
\end{figure*}

\begin{figure}[h!]
\centering
\includegraphics[width=0.49\textwidth]{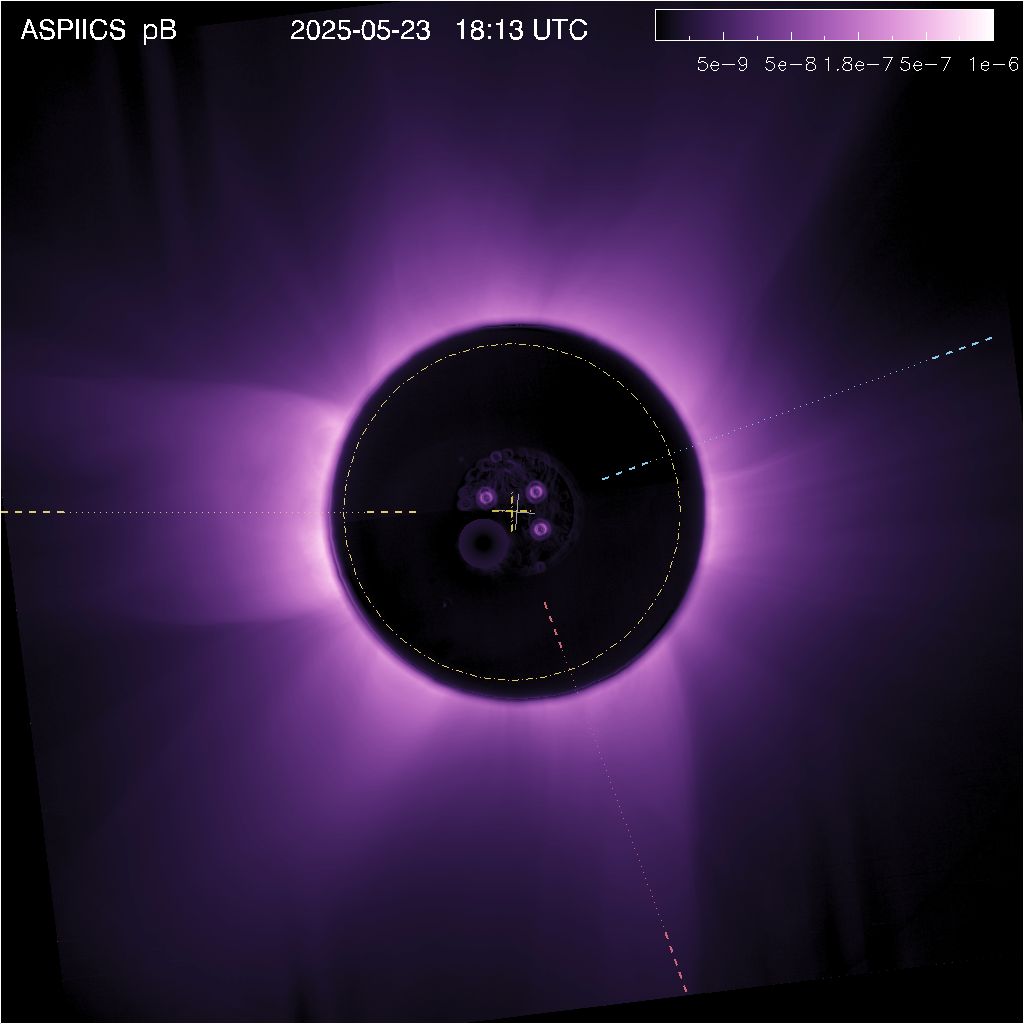}
\caption{Polarized brightness $pB$ of the solar K-corona observed by Proba-3/ASPIICS on 23 May 2025 at 18:13~UTC (orbit 207) in the wide spectral passband centered at 5510~\AA. This high dynamic range image is composed of three images taken with exposure times of 0.2~s, 2~s, and 20~s, each of them made of three polarized images. The internal occulter center is plotted as a gray cross. The position of the center of the Sun and the solar disk are shown as a yellow cross and yellow circle, respectively. Three dotted lines mark the directions (same as in Figure~\ref{fig:radiometry}) along which the polarized brightness of the corona is plotted in Figure~\ref{fig:polar_plots}.}
\label{fig:polar_image}
\end{figure}

\begin{figure}
\centering
\includegraphics[width=0.5\textwidth]{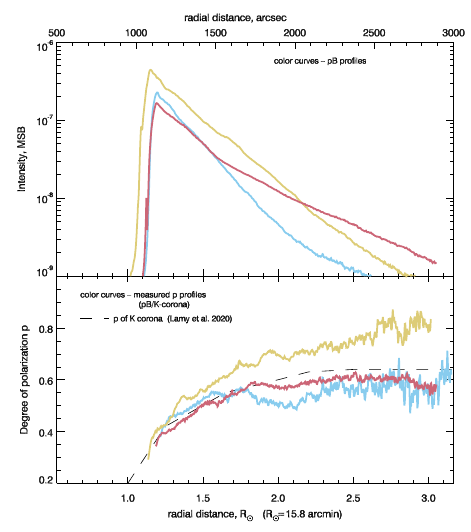}
\caption{Top panel: Profiles of the polarized brightness $pB$ of the K-corona along the dotted lines shown in Figure~\ref{fig:polar_image} are plotted as a function of the radial distance. Bottom panel: corresponding curves of the polarization degree $p$. The dashed curve shows the typical polarization degree of the K-corona of the maximum type obtained by \citet{Lamy2020} based on the classical model of \citet{Baumbach1937}, see also Gibson et al. (2025).}
\label{fig:polar_plots}
\end{figure}

\begin{figure*}[h!]
\centering
\begin{subfigure}[t]{0.5\textwidth}
\centering
\includegraphics[height=7cm]{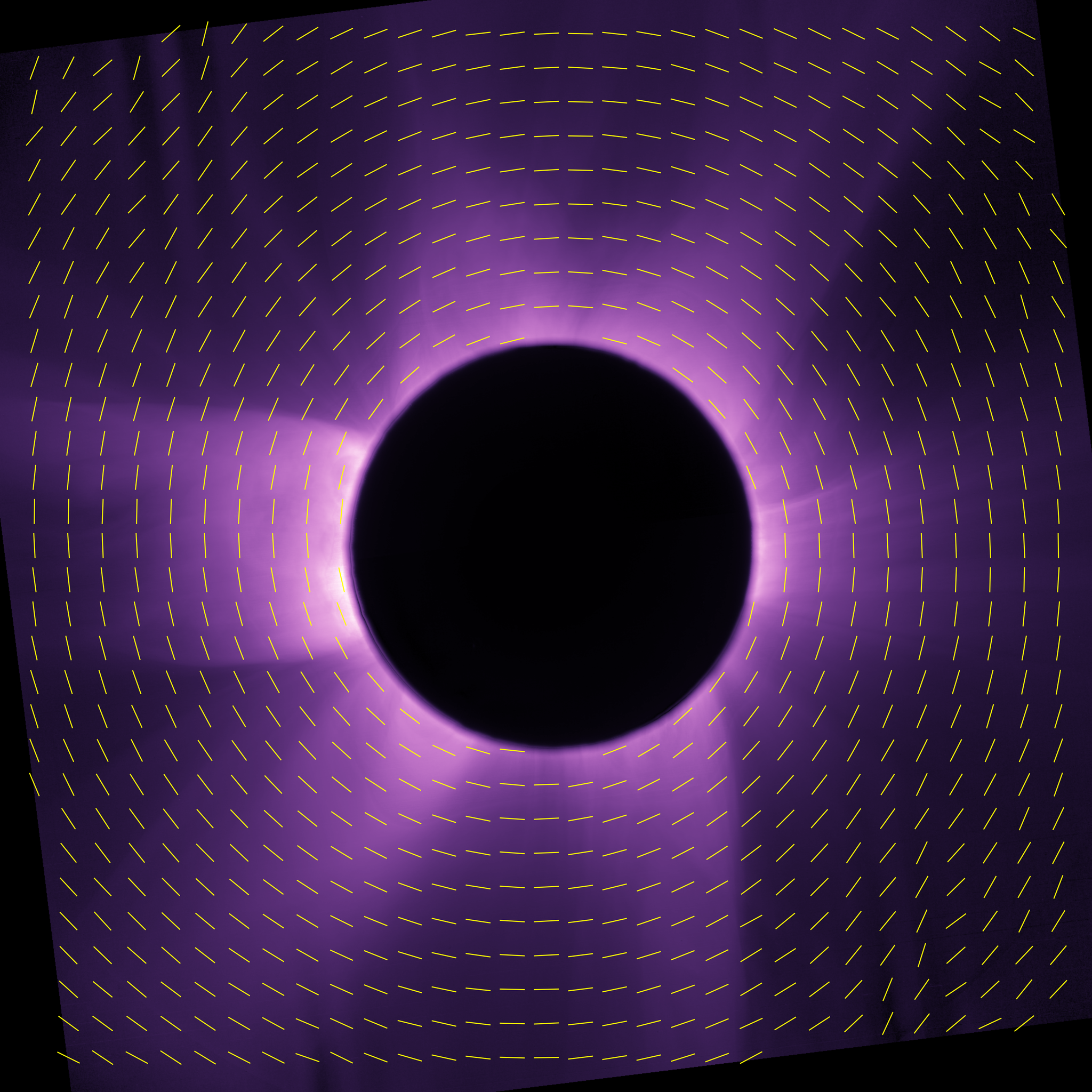}
\end{subfigure}~\begin{subfigure}[t]{0.5\textwidth}
\centering
\includegraphics[height=7cm]{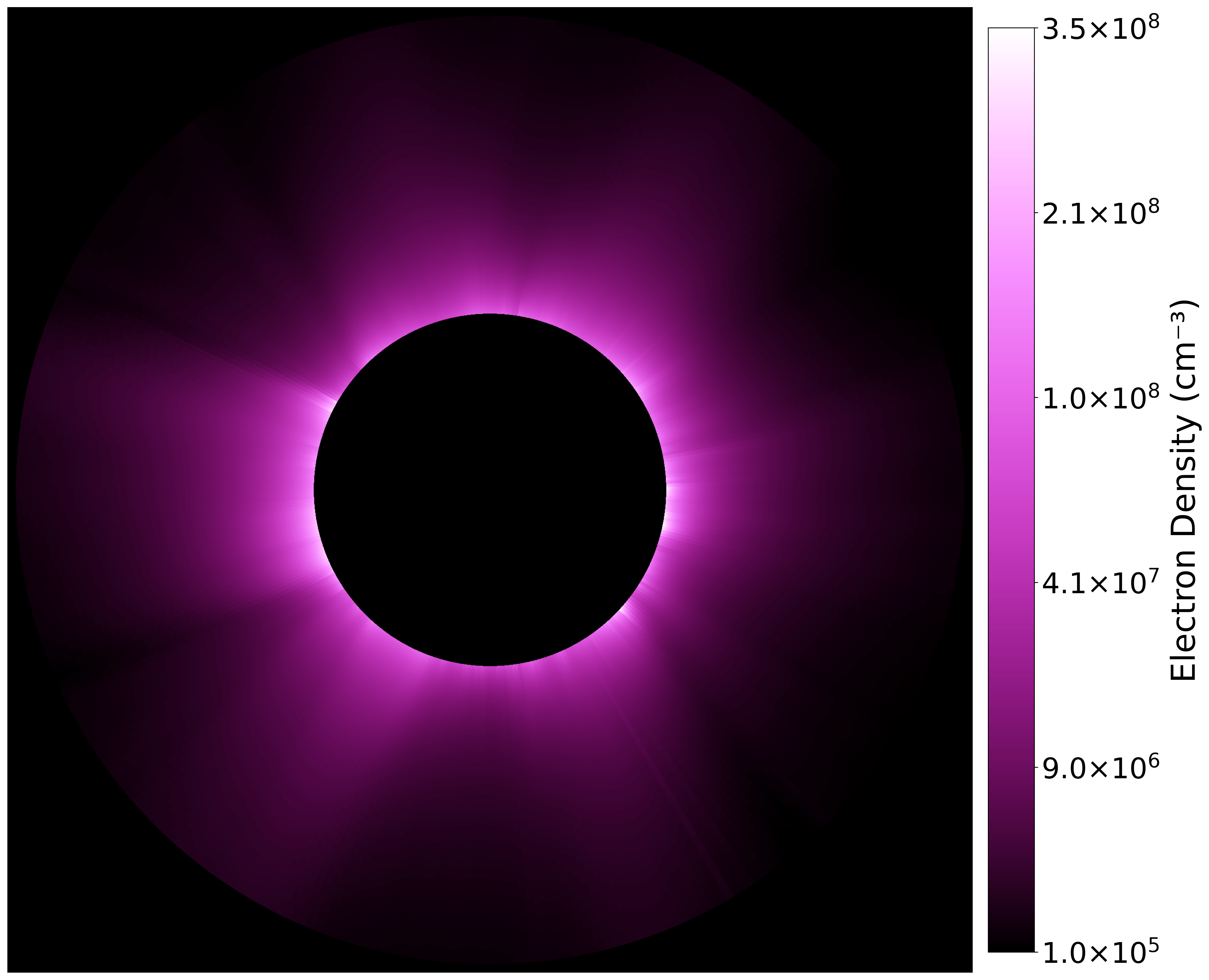}
\end{subfigure}
\caption{Left panel: Polarized brightness $pB$ of the solar K-corona observed by Proba-3/ASPIICS on 23 May 2025 at 18:13~UTC in the wide spectral passband centered at 5510~\AA\ (same image as in Figure~\ref{fig:polar_image}). The bars show the polarization direction. Right panel: Coronal electron number density derived from the polarized brightness measurements (shown in the left panel) using the method by \citet{vandeHulst1950}.}
\label{fig:density}
\end{figure*}

\begin{figure*}[!h]
\sidecaption
\includegraphics[width=12cm]{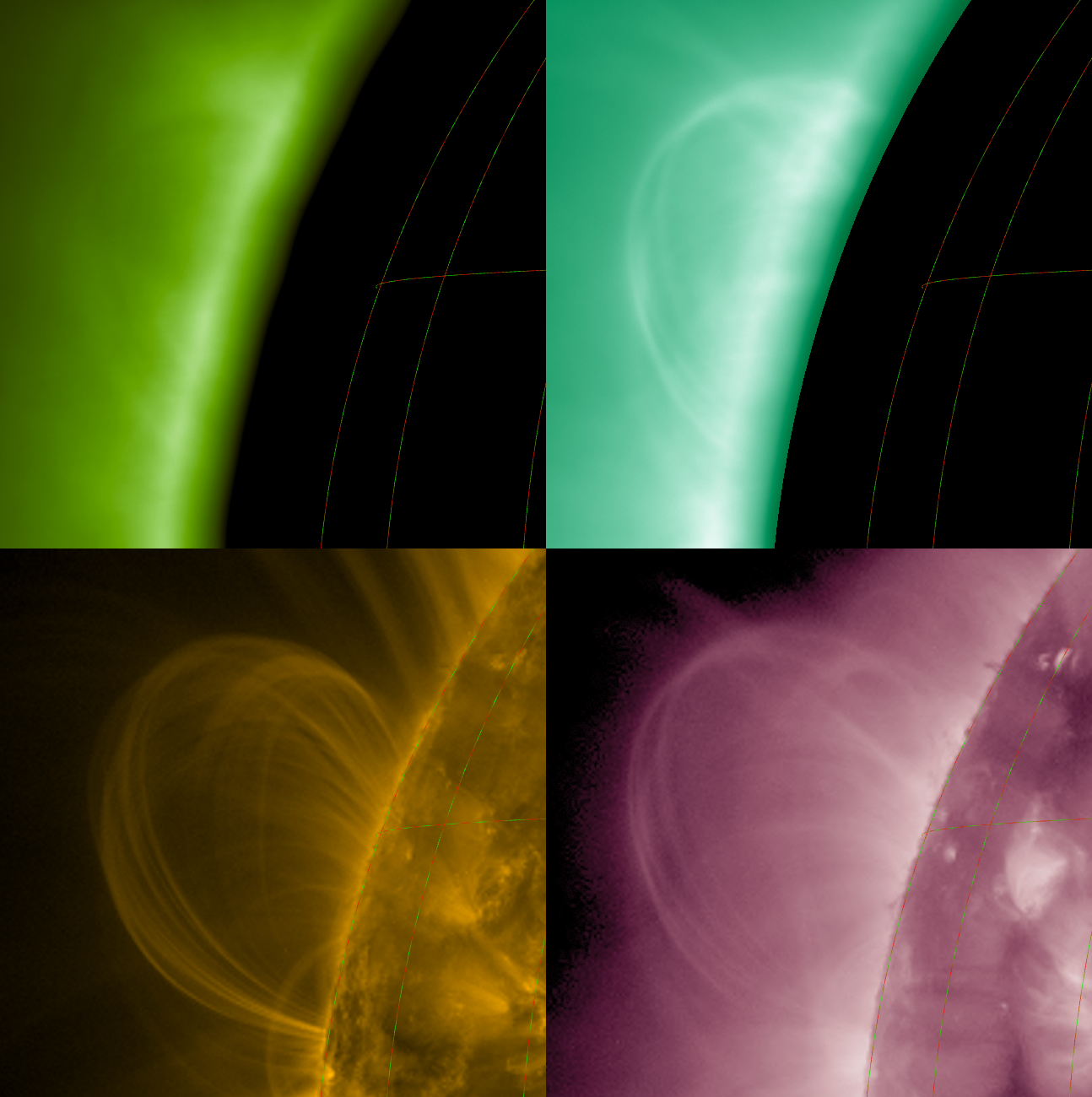}
\caption{The corona above the north-east limb observed on 23 May 2025 by \mbox{Proba-3/ASPIICS} in the wide white-light passband (5510~\AA) at 17:50:32~UTC (top left), by \mbox{Proba-3/ASPIICS} in the Fe~XIV passband (5503~\AA) at 18:03:22~UTC (top right), by SDO/AIA in the Fe~IX passband (171~\AA) at 18:03:09~UTC (bottom left), and by SDO/AIA in the Fe~XIV passband (211~\AA) at 18:03:09~UTC (bottom right). The solar coordinate grid is also shown. The hot 2~MK loops are best seen in the Fe~XIV images taken by ASPIICS and AIA. They are not very well seen in the wide passband of ASPIICS insensitive to coronal temperature. The Fe~IX passband of SDO/AIA is sensitive mostly to the emission of plasma at temperatures around 0.6~MK, so the morphology and contrast of the observed loops are somewhat different in this image.}
\label{fig:fe}
\end{figure*}

\begin{figure}[!h]
\centering
\includegraphics[width=0.49\textwidth]{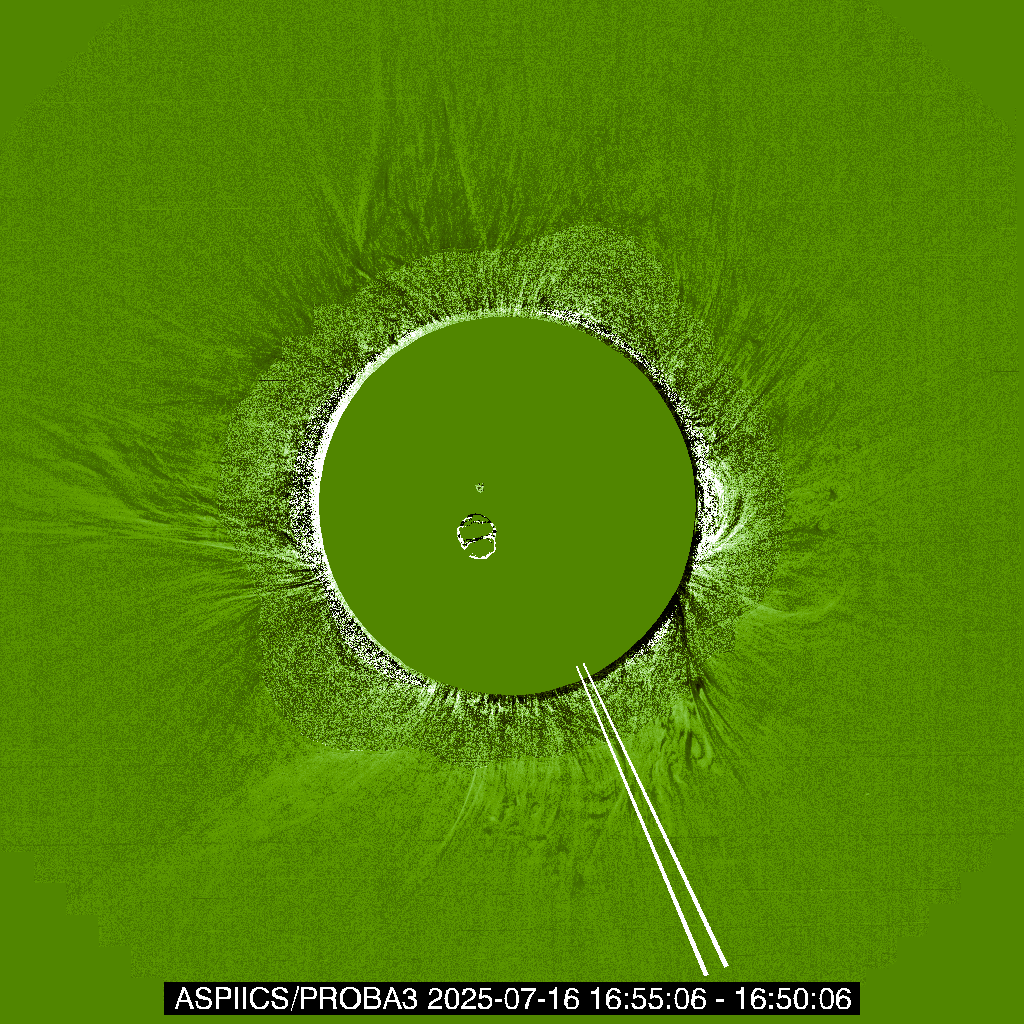}
\caption{Difference image of the corona taken by \mbox{Proba-3/ASPIICS} in wide white-light passband (5510~\AA) on 16 July 2025 at 16:55:06~UTC (orbit 273). A previous image taken at 16:50:06~UTC was subtracted. The 2$^\circ$-wide sector centered at the position angle 204$^\circ$ (measured counterclockwise from solar north) is indicated with two white radial lines. The distance-time map of radial propagation of structures along this sector is shown in Figure~\ref{fig:jmap}. The stitches between individual exposures used to assemble this image, with exposure times of 0.1~s (innermost region), 1~s (middle region), and 10~s (outer region), can be seen. The stitches are not so visible in non-differenced images, see Figures~\ref{fig:radiometry}, \ref{fig:wb_wow} and \ref{fig:polar_image}. The associated movie (Movie 2) is available online.
}
\label{fig:track}
\end{figure}

\begin{figure}[!h]
\centering
\includegraphics[width=0.49\textwidth]{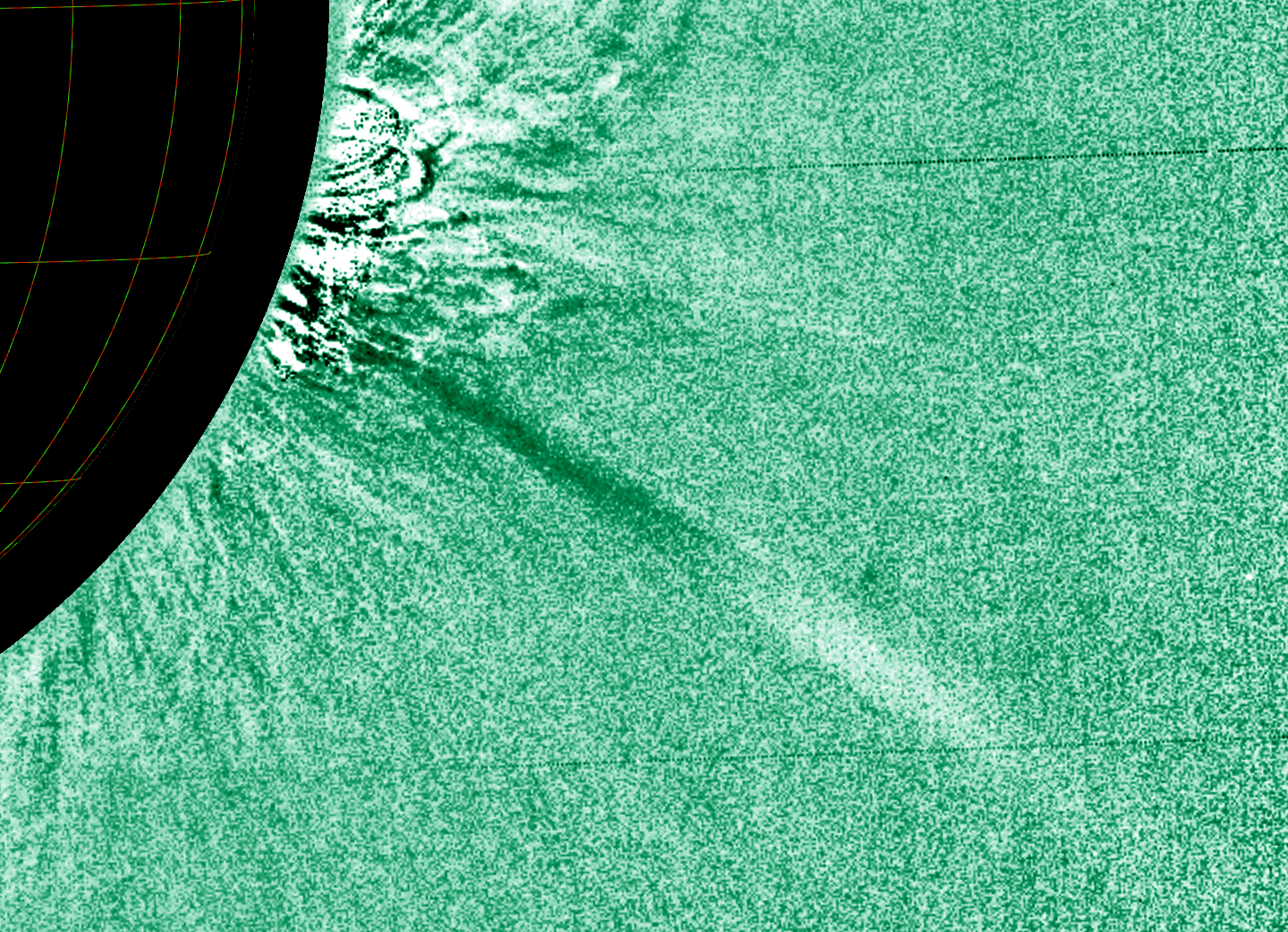}
\caption{A coronal jet observed by Proba-3/ASPIICS in the Fe~XIV (5303~\AA) passband on 16 June 2025 at 09:48:47~UTC (orbit 236). A previous image taken at 09:38:47~UTC was subtracted. The jet first appeared in the ASPIICS field of view at 09:17~UTC and was visible for at least 50 minutes. The jet width was around 2$^\circ$, a typical value for these phenomena \citep{Wang1998}. The associated movie (Movie 1) is available online.  
}
\label{fig:jet}
\end{figure}

\end{appendix}
\end{document}